\documentclass[final,authoryear,1p,times]{elsarticle}

\usepackage{graphicx}

\usepackage{amssymb}
\usepackage{amsmath}
\usepackage{lineno}
\usepackage{numcompress}
\usepackage{color}

\journal{J. Atm. Sol-Ter. Phys.,}

\begin{document}

\begin{frontmatter}



\title{Characteristics of solar diurnal variations: a case study based on records from the ground magnetic station at Vassouras, Brazil}


\author[label1,label2]{Klausner V.}
\author[label1,label3]{Papa A. R. R.}
\author[label2]{Mendes O. Jr.}
\author[label4]{Domingues M. O.}
\author[label5]{Frick P.}

\address[label1]{National Observatory - ON 20921-400, RJ, Brazil}
\address[label2]{INPE/CEA/DGE National Institute for Space Research - INPE 12227-010 S\~ao Jos\'e dos Campos, SP, Brazil}
\address[label3]{Rio de Janeiro State University - UERJ, RJ, Brazil}
\address[label4]{INPE/CTE/LAC National Institute for Space Research - INPE 12227-010 S\~ao Jos\'e dos Campos, SP, Brazil}
\address[label5]{Institute of Continuous Media Mechanics, Perm, Russia}

\begin{abstract}
The horizontal component amplitudes of magnetograms recorded by ground-based observatories of the INTERMAGNET network have been used to analyze the global pattern variance of the solar diurnal variations.
Those kinds of data present gaps in records and consequently we explore them via a time-frequency gapped wavelet algorithm.
We propose a new approach to analyze magnetograms based on scale correlation.
The results show that the magnetic records have a latitudinal dependence affected by the season of year and by the level of solar activity.
We have found a disparity on the latitudinal response at Southern and Northern Hemispheres during solstices, which is expected due to the asymmetry of the Sq field.
On the other hand at equinoxes, records from stations located at approximately the same latitude but at different longitudes presented peculiar dissimilarities.
The achieved results suggest that quiet day patterns and the physical processes involved in their formation are strongly affected by: the conductivity of the E-region, the geomagnetic field intensity and its configuration, and the thermospheric winds.
\end{abstract}


\end{frontmatter}
\linenumbers

\section{Introduction}
\label{Introduction}
The daily variations of the geomagnetic field were discovered by the English researchers \Citet{Graham1724} through the observation of a compass needle motions in 1722.
Since then, it is well known that a typical spectrum of those magnetic variations is composed by a few harmonics of the 24-h period (12, 8 and 6-h).
They are described as ``quiet daily geomagnetic field variations'' - referred to as ``Sq'' for ``solar quiet'' field \citep{Campbell1989}.

The traditional method of calculating the baseline for the quiet day variations is to use the five quietest days for each month for each magnetic observatory.
In this work, we propose a new approach to study quiet periods by eliminating the disturbed days using a multi-scale process.
To accomplish this task, we study the harmonics of the solar diurnal variations using hourly data of the H component and explore it via a gapped wavelet technique based on the continuous wavelet transform \citep[described in][]{Fricketall1997,Fricketall1998}.
The gapped wavelet technique is suitable for analysis of data with gaps.

Originally continuous wavelet transforms (CWT) were applied in geophysics to analyze seismic signals in the pioneer works of \Citet{Morlet1983} and \Citet{GrossmannMorlet1984}.
Nowadays, the use of the wavelet technique has exponentially grown in many different areas \citep{Farge1992}.
A similar trend is noted in the application of the wavelet cross-correlation technique, which has been used by many researchers, see for instance \Citet{Fricketall2001,Oczeretkoetall2006,RehmanSiddiqi2009} and the pioneer work of \Citet{Nesme-Ribesetall1995}.



This work aims mainly to highlight and interpret the solar diurnal variations at a Brazilian station compared to the observations at other twelve magnetic stations reasonably well distributed over the whole Earth's surface. 
By applying gapped wavelet transforms to these signals, we were able to analyze both the frequency content of each signal and the time dependence of that content. 
After computing the wavelet transform, we performed wavelet cross-correlation analysis, which was useful to isolate the period of geomagnetic spectral components in each station and to correlate them as function of scale (pseudo-period or central-period).

The paper is organized as follows: 
Section~\ref{The Physics of the Solar Magnetic Variation} is devoted to explaining the principal mechanisms and aspects of the solar magnetic variations. 
Section~\ref{Magnetic Data}, the analyzed period and data are presented. 
Section~\ref{Methodology} describes, divided in subsections, the used methodology: 
Section~\ref{Continuous Wavelet Transform} presents a brief description of continuous wavelet transforms, 
Section~\ref{Gapped Wavelet Analysis} is devoted to introduce the gapped wavelet analysis,
and Section~\ref{Wavelet Cross-correlation Analysis} to establish the wavelet cross-correlations and to explain how they can be quantified. 
Section~\ref{Results and Discussion}, the results are discussed and,
Section \ref{Summary} brings the conclusions of this work.
The paper also includes two appendixes which we present some CWT application and mathematical details concerning to this work.

\section{The Physics of the Solar Magnetic Variation}
\label{The Physics of the Solar Magnetic Variation}

The major driving force for quiet day field changes seems to arise from the dynamo-current process in the ionospheric E region between 90 and 130 km \citep{Stewart1882}. 
Tidal winds move the ions across the Earth's magnetic field producing electro-magnetic forces (emfs). 
Those emfs drive electric currents in the conducting E region which give rise to daily variations in the magnetic field measured at the ground level \citep[for details see][]{ChapmanBartels1940}. 
Through these mechanisms, two vortices of currents are induced, one in the northern Hemisphere (clockwise) and another in the southern Hemisphere (counterclockwise).
At the same time, a strong eastward electric jet is formed throughout the equatorial region.

There are three factors that affect the dynamo process: the ionospheric wind, the ionospheric conductivity and the geomagnetic field configuration.
The wind and the conductivity vary seasonally due to their dependence on the solar zenith angle \citep{Campbell1989}.
\Citet{Zhaoetall2008} concluded that the correlation between the Sq amplitude and solar zenith was higher in high latitude than in low latitude regions due to the effect of the prenoon-postnoon asymmetry of Sq.

The Sq field variation has a main spatial dependence on latitude and is affected by other factors including epoch of the year and level of solar activity.
\Citet{ChapmanStagg1929} observed two kinds of regular changes in the solar diurnal magnetic variations on quiet days: annual variation, that affects both the type and the amplitude during each year; and solar activity variation, that affects fundamentally the amplitude along each sunspot cycle.
\Citet{Takeda1999} estimated that the intensity of the Sq currents in high solar activity was about twice larger than in low solar activity.

Many other researchers studied the variations of the Sq, including \Citet{Hibberd1985} that examined the annual, semi-annual and even the whole solar cycle.
\Citet{Stening1971} examined seasonal variations and longitudinal inequalities of the electrostatic-field in the ionosphere by looking at its electric conductivity and the Earth's main magnetic field.
\Citet{Takeda2002} showed solar activity dependence of the Sq amplitude, and explained this effect through the ionospheric conductivity. 
\Citet{Takeda2002} also compared the amplitude of the Sq for the same value of conductivity.
The seasonal variation is seemingly due to differences in neutral winds or to the magnetic effect of the field-aligned current (FAC) flowing between the two Hemispheres generated by the asymmetry in the dynamo action. 
The FACs are controlled by interplanetary magnetic fields (IMF) and its electric fields can directly penetrate to the equatorial ionosphere \citep{Sastri1988}.

Some evidence of the influence of oceanic tides on the magnetic daily variation has been obtained by \Citet{LarsenCox1966}. 
They found small semidiurnal variations of the Z component at a coastal site (Cambria, California) and at two island stations (Honolulu and San Miguel) that could not be explained by the atmospheric tidal theory.
They suggested that these variations must be due predominantly to oceanic tides. 
It is important to mention here that the conductivity of the ocean does not vary significantly with time, unlike the ionospheric conductivity.
As a consequence, the seasonal variation of the oceanic contribution is expected to be smaller than the ionospheric contribution \citep{Cuetoetall2003}.

Magnetic stations on islands or coastlines are fully exposed to anomalous effects of the ocean \citep{Schmucker1999a, Schmucker1999b}.
The oceans, as well as the ionosphere, constitute electrical conductors and they are subject to tidal motions.
As a result of the tidal flow of the ocean water, we might expect the effects of some kind of dynamo \citep{Parkinson1983}. 
The electric current system induced by the oceanic tides due to the drifting motion produces daily variation on magnetic stations located at oceanic islands and nearby shores. 
Caution must be taken when interpreting the solar magnetic variations at a particular station, if the purpose is to obtain a world-wide analysis of the Sq field. 

The total variation measured on the ground consists both of external (ionospheric current) and internal (induced Earth current) contributions \citep{ForbesLindzen1976}.
The pattern of the conductivity of surfaces layers of the Earth will introduce a corresponding small scale pattern into the distribution of the induced currents.
At some stations, an accurate indication of the average induced current system is difficult to determine, consequently its effects cannot be fully distinguished \citep{Price1969}.

In his review, \Citet{Price1969} suggested that the interpretation of field variations at a particular observatory must be done carefully.
Although the ionosphere current system may have a fairly simple world-wide pattern, the relatively small variation of the conductivity distribution of the Earth's surface will introduce a corresponding small variation into the distribution of the induced currents.
However, the external origin contribution of the Sq field is about 2.5 times that of the internal origin \citep{MatsushitaMaeda1965}. 
Therefore, in this work, we have disregarded the influence of the internal contributions on the analysis of the diurnal geomagnetic variations.

\section{Magnetic Data}
\label{Magnetic Data}

In this section, we first describe the data used to study the quiet day variations. We also discuss the considerations used in data treatment for the different magnetic stations.

\subsection{Dataset}
\label{Dataset}


Our primary interest is to correlate the response of the geomagnetic field at Vassouras to the other twelve previously chosen magnetic stations. 
The Vassouras Magnetic Observatory, is located in Rio de Janeiro, Brazil,  under the South Atlantic Magnetic Anomaly (SAMA) influence.
It has been active since 1915 and is a member of the INTERMAGNET program.
One of the peculiarities of VSS is its location, at low latitude, where the H component is essentially the same as the total geomagnetic field. 
In forthcoming years, a Brazilian network of magnetometers will be implemented and VSS could be used as reference.

To fulfill our purpose, we use hourly mean value series of the H geomagnetic component. 
Some magnetic stations have available the X component, then we convert the X component of XYZ system to the H component of the HDZ system of vector representation of the Earth's magnetic field \citep[as described in][]{Campbell1989}.
We use $ X=\,H\,\cos{\left(D\right)}$, where X is the vertical component in the XYZ system, H is the horizontal magnitude and D is the angular direction of the horizontal component from the geographic north (declination). 
In principle, this system conversion does not affect our results because we only use magnetic stations of low- and mid-latitudes and we are interested in the magnetic variations.

The magnetic stations that use the XYZ system are AMS, ASP, BEL and CLF.
The conversion of systems was performed for BOU, CMO, EYR, HON and SJG for the data between 1999 and 2003 and in BMT and KAK for the data between 1999 and 2004.

\begin{figure}[ht]
  \centering
    \includegraphics[width=1.0\textwidth]{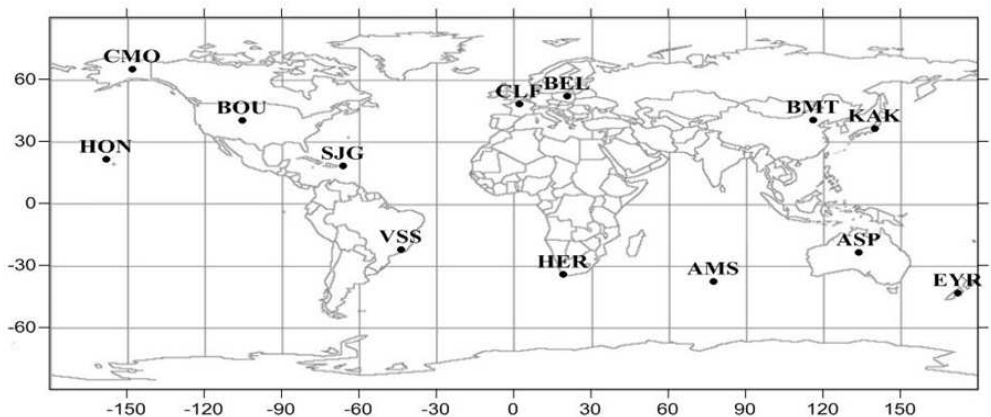}
  \caption{Geographical localization of the stations used in this work and their respective IAGA code.}
  \label{fig:MapStations}
\end{figure}

The distribution of the magnetic stations, with their IAGA code, is given in Fig.~\ref{fig:MapStations}. 
The corresponding codes and locations are given in Table~\ref{table:ABBcode} which the sequence is organized by the geographical latitude of the stations.
This work relies on data collections provided by the INTERMAGNET programme (http://www.intermagnet.org).

\begin{table}[ht]
 \caption{INTERMAGNET network of geomagnetic stations used in this study.}
\centering
\begin{tabular}{l l r r r r}
\hline
Station& IAGA code & \multicolumn{2}{c}{Geographic coord.} & \multicolumn{2}{c}{Geomagnetic coord.}\\
\cline{2-6}
       &   & Lat.($^o$) & Long.($^o$) & Lat.($^o$) & Long.($^o$)  \\[0.5ex]
\hline
Eyrewell (New Zealand) &EYR     &-43.42     &172.35        &-46.79     &-106.06   \\
Martin de Vivies (France) &AMS     &-37.83     &77.56         &-46.07   &144.94  \\
Hermanus (South Africa) &HER     &-34.41     &19.23         &-33.89     &84.68   \\
Alice Springs (Australia) &ASP     &-23.76     &133.88        &-32.50   &-151.45   \\
Vassouras (Brazil) &VSS     & -22.40    & -43.65       &-13.43     &27.06 \\
San Juan (Puerto Rico) &SJG     &18.12      &-66.15        &27.93      &6.53  \\
Honolulu (United States) &HON     &21.32      &-158.00       &21.59      &-89.70   \\
Kakioka (Japan) &KAK     & 36.23     & 140.18       & 27.46     &-150.78\\
Boulder (United States) &BOU     & 40.13     & -105.23      & 48.05    &-38.67   \\
Beijing (China) &BMT     &40.30      &116.20        &30.22     &-172.55   \\
Chambon la Foret (France) &CLF     &48.02       &2.26         &49.56      &85.72   \\
Belsk (Poland) &BEL     &51.83      &20.80         &50.05    &105.18   \\
College (United States) &CMO     &64.87      &-147.86       &65.36      &-97.23   \\[1ex]
\hline
\end{tabular}
\\Source: http://wdc.kugi.kyoto-u.ac.jp/igrf/gggm/index.html (2010)
\label{table:ABBcode}
\end{table}

We choose the period of geomagnetically quiet days to study seasonal and solar activity variation. 
The data interval used in this work is from 1999 to 2007, almost the whole solar cycle 23 using the available dataset. 
In this data interval, the higher solar activity occurred between 2000 and 2001; and the lower solar activity, between 2005 and 2007.
For seasonal variation analysis, one year is divided into two seasons: solstices (June and December) and equinoxes (March and September).

As a geomagnetic disturbance index, the Kp index has been chosen to distinguish the disturbed days ({$Kp>3+$}) from the quiet days ({$Kp\le 3+$}).
The global Kp index is a number from 0 to 9 obtained as the mean value of the disturbance levels within 3-h interval observed at 13 subauroral magnetic stations \citep[see][]{Bartels1957}.
In our approach, the disturbed periods are eliminated and considered as gaps.
Therefore, in principle only the magnetic effects of the lunar and solar contribution remained in the magnetic records.

\subsection{Data Treatment Considerations}
\label{Data Treatment Considerations}

As previously mentioned, we are focusing the study of VSS responses to Sq variations, particularly, its diurnal and semidiurnal period spectral components.
As VSS is located at low latitude, the majority of magnetic stations used here are spread between low- and mid-latitudes.
We only use one magnetic station (CMO) located at high latitude for comparison.
In the polar region, disturbance events have been included because there were not enough truly quiet days with our Kp selection ($Kp\le 3+$).
Above $60^o$, magnetospheric processes may completely dominate the magnetic recordings what impedes the Sq observation \citep{Campbell1989}.
At low- and mid-latitude, the magnetograms should be not seriously affected by auroral electrojet and the Sq currents would correspond to regularly recurring phenomena \citep{Price1969}.

The Sq variations were greatly explored over the last decades, as well as their spatial dependence on latitude and others factors, including epoch of the year and level of solar activity \citep{Price1969, Stening1971, Hibberd1985, Campbell1997, SagerHuang2002, Takeda2002}. 

We take into consideration the two well known facts about the Sq field cited by \Citet{Price1969}: 
\begin{itemize}
\item[1.]
It is largely a local time field that can be roughly represented by a current fixed to the Sun, then we use the data at Universal Time ($\mathrm{UT}$) to observe the global variance of the Sq field.
\end{itemize}
\begin{itemize}
\item[2.]
 It is influenced by the Earth's main field throughout the latitudes between 60$^o$N and 60$^o$S, then we do not use magnetic stations of high latitudes (auroral zone), except CMO. 
\end{itemize}

On one hand, at high and polar latitudes, the geomagnetic field lines are aligned almost vertically then the ionospheric currents are joined with the field-aligned currents (FAC), and the electrodynamics is dominated by the influence of the solar-wind-magnetosphere interaction processes \citep{Yomoto2006}. 
On the other hand, the ionospheric current at middle and low latitude is generated by the influence of tidal winds, \textit{i.e.}, ionospheric wind dynamo \citep{Sastri1988}.

An adequate knowledge of the daily variation field and a full understanding of the associated phenomena can only come from extensive and detailed analysis of the mean hourly values of the magnetic elements at many stations \citep{Price1969}. 
Therefore, the availability of these data in the World Data Centers that collect and distribute these data is of enormous help.

\section{Methodology}
\label{Methodology}

The method used in this study is based on continuous wavelet transforms. 
In this section, we first introduce the concepts of the CWT and its properties.
Following, we introduce the gapped wavelet analysis and emphasize its improvements over the CWT.
After that, we describe how we extract the information of the observed data using wavelet cross-correlation analysis.

\subsection{Continuous Wavelet Transform}
\label{Continuous Wavelet Transform}

The wavelet transform was introduced at the beginning of the 80s of the XX century by \Citet{Morlet1983}, who used it to study seismic data. \Citet{GrossmannMorlet1984} improved the windowed Fourier transform and they constructed the continuous wavelet transform (CWT). 
The idea was to change the width of the window function accordingly to the frequency of the signal being considered. 
The CWT is an integral transform, and it yields an affine invariant time-frequency representation. 
A wavelet transform can measure the time evolution of the frequency transients within the signal. 
In this transform, we use as basis a set of functions called wavelets.

Formally, in order to be called wavelet, a function $\psi$ must satisfy some conditions. 
First, the admissibility condition which is usually related to the mean zero of the wavelet function,

\begin{equation}
\label{admiss}
   \int \psi (t)\,dt=0.
    \end{equation}

Second, the wavelet function should be localized with compact or effective compact support.

The CWT of a time series $ f $ is defined by the integral transform,

\begin{equation}
 W(a,b) =\, \int f(t) \, \frac{1}{\sqrt{a}}\, \psi^{*} \left(\frac{t-b}{a}\right)\, dt, \; \; \; \: \: a>0,\: a \in \mathbb{R},
\end{equation}
where $*$ represent the complex conjugate and the pre-factor $\left| a \right| ^{\frac{1}{2}}$ is introduced in order to guarantee that all the scaled functions  $ \left| a \right| ^{\frac{1}{2}}\,\psi (\frac{t}{a}) $, have the same energy in $\mathbb{L}^2(\mathbb{R})$ sense.
This function $W(a,b)$ represents the wavelet coefficients that is a function of both time and frequency (time $b$ and scale $a$). 


There are many possible kinds of wavelets.
The choice of wavelets depends both on the data and the analysis objectives \citep[][and references therein]{Domingues2005}.
Therefore, there is no best wavelet for signal analysis in general.
In this case, we use the Morlet wavelet function that represents well the signals we are analyzing.

It is possible to analyze a signal in a time-scale plane. 
In the wavelet analysis it is called the wavelet scalogram.
In analogy with the Fourier analysis, the square modulus of the wavelet coefficient $|W(a,b)|^2$ is used to provide the energy distribution in the time-scale plane. 
In the wavelet analysis, we can also explore the central frequencies $\varepsilon_{\psi}$ (or central periods $\frac{1}{\varepsilon_{\psi}}$ ) of the time series through the global wavelet spectrum, also called pseudo-frequencies $\varepsilon^{~}_{a}$ (or pseudo-periods ),

\begin{equation}
\label{eq:pseudofreq}
 \varepsilon^{~}_{a}=\frac{\varepsilon_{\psi}}{a\,\Delta t.}
\end{equation}

\noindent
where $\Delta t$ is the sampling period \citep{Abry1997}.

It helps to understand the behavior of the energy at a certain scale. 

\subsection{Gapped Wavelet Analysis}
\label{Gapped Wavelet Analysis}

Magnetograms are finite length observational data series and may contain gaps of various sizes. 
To reduce gap problems, we use the gapped wavelet analysis — a technique introduced by \Citet{Fricketall1997} and afterwards improved in \Citet{Fricketall1998}.
In this transform, the admissibility condition is broken when the wavelet overlaps data gaps.
The leading idea of the gapped technique is to restore the admissibility condition by repairing in some way the wavelet itself.

Considering the function $f(t)$ is only known in some intervals of time and it can be rewritten as

\begin{equation}
 f'(t)=f(t)\,G(t)
\end{equation}
\noindent
where $G(t)$ is the function of data gaps, which is equal to $1$ if the signal is registered and is equal to zero otherwise, therefore, the analyzing wavelet became

\begin{equation}
 \psi _{a,b}' (t)=\,\frac{1}{\sqrt{a}}\, \psi \left(\frac{t-b}{a}\right)\,G(t).
\end{equation}

Near a gap, the function $\psi'$ is used instead of the base wavelet $\psi$, and consequently, $\psi'$ will no longer satisfy the admissibility condition (Equation~\ref{admiss}).
The function $G$ transfers the gap problem from the signal $f(t)$ to the broken wavelet $\psi'$, which will be replaced by an adaptive wavelet $\widetilde{\psi}$ in order to satisfy the admissibility condition.

To restore the admissibility condition, the analyzing wavelet $\psi$ can be considered to be in the form

\begin{equation}
 \psi (t)= \,h(t)\:\varphi(t),
\end{equation}
\noindent
where $h(t)$ is the oscillatory part and $\varphi(t)$ is the envelope.

In this analysis, we use the Morlet wavelet,

\begin{equation}
 \psi(t)=\, \exp\left(\dfrac{-t^{2}}{2\:\sigma^{2}}\right)\:\exp(\imath \,\omega_{0}\,t),
\end{equation}
with $\imath=\sqrt{-1}$, $\omega_{0}=6$ and $\sigma=1$, $\sigma$ is the time resolution parameter.
Using Morlet wavelet with $\omega_{0}=\,6$ gives a value of $\varepsilon_{\psi}=0.9709$ and Fourier period, $T$, equal to $T=1.03\, a$, where $a$ is the scale.
It indicates, that for the Morlet wavelet, the scale is approximately equal to the Fourier period \citep[see][for more details]{Farge1992, Abry1997,TorrenceCompo1998}. 

The adjustable parameter $\sigma$ gives the optimal time-frequency resolution.
Small values of $\sigma$ give better time resolution, while large values improve frequency resolution.
In the gapped wavelet case, the problem of the admissibility condition breaks down when $\sigma$ is below 1.
It must be avoided.
However, a wide range of $\sigma$ can be used.
The choice of $\sigma$ is highly restricted by the admissibility condition, because there are not enough oscillations to give a zero in average.
However, the gapped technique corrects the wavelet in this case as well and nullifies the mean.
Thus this problem is completely resolved by the gapped technique \citep{Soonetall1999}.
For more details see \ref{Appendice A}.

Following \Citet{Fricketall1997}, we use Morlet wavelet where

\begin{equation}
 h(t)=\,\exp(\imath \,\omega_{0}\,t),
\end{equation}
\begin{equation}
 \varphi(t)=\,\exp\left(\frac{-t^{2}}{2\:\sigma^{2}}\right).
\end{equation}

When the wavelet is disturbed by the gap, we can restore the admissibility condition by including a function $\mathsf{K(a,b)}$ in the oscillatory part of the wavelet,

\begin{equation}\label{eq:phitil}
 \widetilde{\psi}(t,b,a)=\,{\left[ h \left(\frac{t-b}{a}\right)\, -\,\mathsf{K}(a,b) \right]}\:\varphi\left(\frac{t-b}{a}\right)\,G(t)
\end{equation}

\noindent and requiring,

\begin{equation}\label{eq:admisstil}
   \int  \widetilde{\psi} (t)\,dt=0.
    \end{equation}

The introduced function $\mathsf{K}(a,b)$ can be determined for each scale $a$ and position $b$ from (\ref{eq:phitil}) and (\ref{eq:admisstil}), and could be obtained as

\begin{equation}
 \mathsf{K}(a,b) =\,{\left[ \int \varphi \left(\frac{t-b}{a}\right)\,G(t)\,dt\right]}^{-1}\:{\left[ \int h \left(\frac{t-b}{a}\right)\,\varphi \left(\frac{t-b}{a}\right)\,G(t)\,dt\right]}.
\end{equation}

It was shown that this technique not only suppresses the noise caused by the gaps and boundaries, but improves the accuracy of frequency determination of short or strongly gapped signals.

In summary, the advantages of the CWT using gapped wavelet in comparison with the use of the traditional CWT are: \\

\begin{enumerate}
 \item The gapped wavelet technique helps to reduce the effects of two problems of the detection of periodicities in times series: the presence of gaps in time series and boundary effects due to the finite length of time series.
\item The method involves a correction of the analyzing wavelet to fulfill the admissibility condition and is independent of a particular choice of analyzing wavelet.
\item This technique not only suppresses the low and high noise frequencies, but it is also better at estimating the frequency of the signal.
\end{enumerate}

In \ref{Appendice B}, we present an example of the gapped wavelet use in a synthetic signal.

\subsection{Wavelet Cross-correlation Analysis}
\label{Wavelet Cross-correlation Analysis}

The approach of this work is to use the wavelet cross-correlation, $\mathcal{C}(a)$, to study the correlation between a pair of magnetic data from different stations as a function of scale (see \Citet{Nesme-Ribesetall1995} and \Citet{Fricketall2001} for more mathematical details):

\begin{equation}
\label{eq:corr}
 \mathcal{C}(a)=\, \frac{\int \mathcal{W}_{1}(a,t)\,\mathcal{W}_{2}^{*}(a,t)\,dt}{\left(\int \mathcal{W}_{1}(a,t)^{2}\,dt\:\int \mathcal{W}_{2}(a,t)^{2}\,dt\right)^\frac{1}{2}}
\end{equation}
where $\mathcal{W}_{i}(a,t)=\,\left|W_{i}(a,t)\right|\,-\,\overline{\left|W_{i}(a,t)\right|}$, $W_{i}$ are the wavelet coefficients and $\overline{W_{i}}$ is the arithmetic mean in time for $i=1$ or $2$.
We can also obtain the determination coefficient, $D(a)$, as function of the scale $a$. 
It is defined as the square of the correlation coefficient,
 
\begin{equation}
\label{eq:det}
D(a)=\mathcal{C}(a)^2.
\end{equation}
 
One of the reasons to use the coefficient of determination instead of the correlation is to compute the statistics in order to determine the size or magnitude of the relation between two variables. 
It is interpreted as the percentage of variability of the response variable explained by the regression model. 
The correlation coefficient measures linear association.
Though in space geophysics both the determination and correlation coefficient are used \citep{Reiff1983}. 
In our case, we prefer to use the determination coefficient due to its interpretation of linear regression.

As mentioned above, in the wavelet representation, each scale ($a$) explains part of the distribution of energy of the whole signal through time.
The wavelet cross-correlation allows us to check the interaction between two sets of data for each considered scale. 
The scales are chosen in such a way that they make possible to characterize the dominating periods in the geomagnetic data spectrum.
In our case, we choose the scales which correspond to the periods of $24$ and $12$ hours, related to the diurnal magnetic variations.

\section{Results and Discussion}
\label{Results and Discussion}

\begin{figure}[ht]
  \centering
    \includegraphics[width=1.0\textwidth]{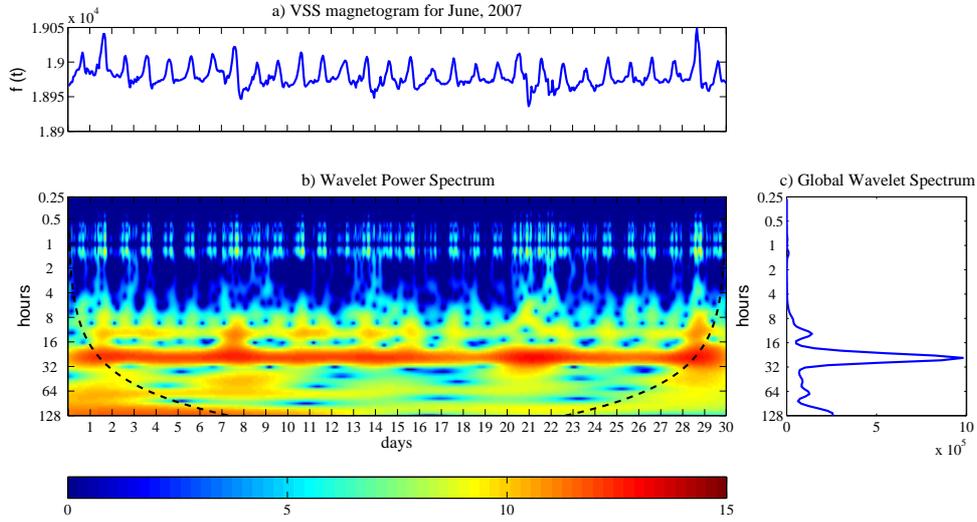}
  \caption{Example of CWT application to a real signal: (a) H component of VSS at June, 2007 used for the wavelet analysis, (b) the scalogram using Morlet wavelet, logarithmic scaled representing $\log2{(|W(a,b)|)}$ and (c) the global wavelet spectrum.}
  \label{fig:VSS2007Jun}
\end{figure}

\begin{figure}[ht]
  \centering
    \includegraphics[width=1.0\textwidth]{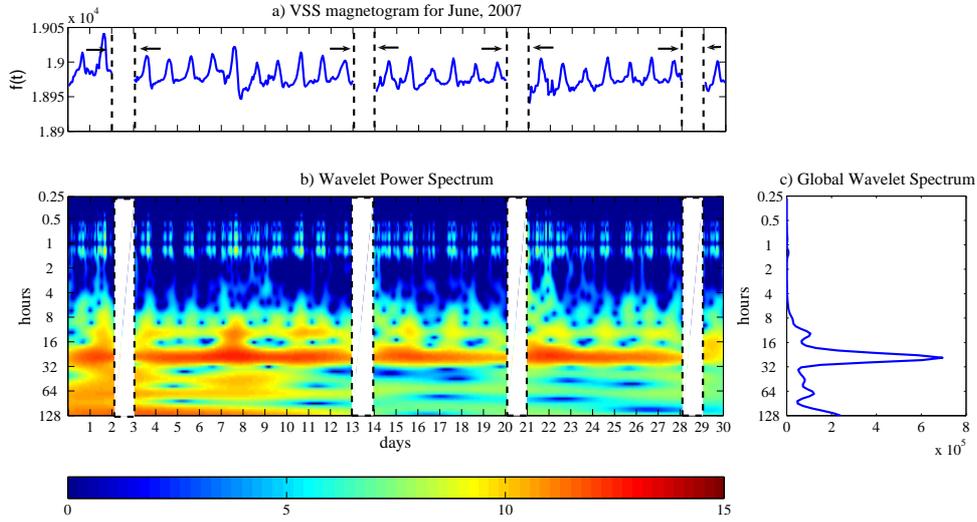}
  \caption{Example of application of gapped wavelets to a real signal with added gaps: (a) H component of VSS at June, 2007 with additional gaps due to the disturbed days; (b) the local wavelet power spectrum using Morlet wavelet, logarithmic scaled representing $\log2{(|W(a,b)|)}$ and (c) the global wavelet spectrum.}
  \label{fig:VSS2007JunGap}
\end{figure}

Fig.~\ref{fig:VSS2007Jun} and Fig.~\ref{fig:VSS2007JunGap} show the interpretation of the CWT and the CWT using gapped wavelet techniques, with the purpose highlighting the reasons to use gapped wavelet techniques over the CWT to study quiet days.
Fig.~\ref{fig:VSS2007Jun} shows an example of CWT applied to a real signal, in this case, a magnetogram data.
In this analysis, as function of time, $f(t)$, we use the H component obtained in June, 2007 at VSS station. 
Fig.~\ref{fig:VSS2007Jun} shows: (a) the H component, (b) the wavelet square modulus (scalogram) and (c) the global wavelet spectrum (total energy in each scale).
In the scalogram, areas of stronger wavelet power are shown in dark red on a plot of time (horizontally) and time scale (vertically).
The areas of low wavelet power are shown in dark blue.
It is possible to notice a maximum of wavelet power on the scalogram and, also on the global wavelet spectrum, at the time scale corresponding to 24-h period, and a less pronounced second peak at the time scale corresponding to 12-h period.
The scalogram shows also the cone of influence, region where edge effects become important and the wavelet coefficients are not reliable.

Magnetograms may contain gaps of various sizes.
Here, we also added gaps due to the removal of the disturbed days.
These disturbed days were removed before the gapped wavelet transform was performed.
In our case, the gapped wavelet technique helps to reduce the effects due to the presence of gaps and also boundary effects due to the finite length of data.
The gapped wavelet (used and validated on the synthetic signal - \ref{Appendice B}) was then applied in real data as shown in Fig.~\ref{fig:VSS2007JunGap}. 
One might notice that Fig.~\ref{fig:VSS2007JunGap} is very similar to Fig.~\ref{fig:VSS2007Jun}, the difference is that now we introduced gaps due to disturbed days.
As our interest is only to study the quiet day variations, we consider the disturbed days (where $Kp>3+$) as gaps. 
Even with the additional gaps, the global wavelet spectrum, Fig.~\ref{fig:VSS2007JunGap}(c), still shows a pronounced increase of energy in the 24-h and a less dominant in the 12-h period.
In the scalogram, the 24-h period shown in dark red is clearly visible and the 12-h period in the reddish color appears less dominantly compared to the 24-h.
Also, it is not necessary to include the cone of influence, because the gapped wavelet technique not only suppresses the boundary and gaps in both high and low frequencies, but also  estimates better the frequency of the signal (see \Citet{Fricketall1998}).
Comparing Fig.~\ref{fig:VSS2007Jun} and Fig.~\ref{fig:VSS2007JunGap}, there is not a significant difference in the common intervals of the CWT and gapped scalograms.
The inclusion of gaps in the data due to the removal of the disturbed days does not affect the analysis of the diurnal and semidiurnal variations.

With the gapped wavelet coefficients in hand, we exclude the coefficients that corresponded to the gaps areas in the scalograms.
The excluded areas are presented as white rectangles in Fig.~\ref{fig:VSS2007JunGap}.
It is an artificial lack of data that follows the methodology established in \ref{Appendice A} (which it has purpose to highlight the gapped wavelet property of the estimating frequency of the signal in the gaps regions similar to an interpolation method), so Fig.~\ref{fig:VSS2007JunGap} highlights the wavelet coefficients excluded before performing the cross-correlation wavelet analysis.
Because we are only interested in the quiet-days behavior analysis.

This methodology structures a new way of dealing with daily magnetic variation for quiet periods.

The same procedure of excluding the disturbed days and consider them as gaps is performed for the whole dataset used in this study.
Also, after the gapped wavelet transform, we have also excluded all the wavelet coefficients that corresponded to these gaps in the scalograms.
Therefore, the cross-correlation wavelet analysis were done only using the wavelet coefficients that corresponded to the quiet days.

After excluding all the wavelet coefficients that correspond to gaps, we have calculated the cross-correlation functions $C \left( a\right)$, for each pair formed by Vassouras and each one of the twelve chosen magnetic stations (see equation~\ref{eq:corr} for mathematical details). 
This procedure was done for the years 1999 to 2007 during the equinoxes and solstices.
These calculations were performed in order to understand the global response of the Sq variation at different locations.

\begin{figure}[ht]
  \centering
    \includegraphics[width=1.0\textwidth]{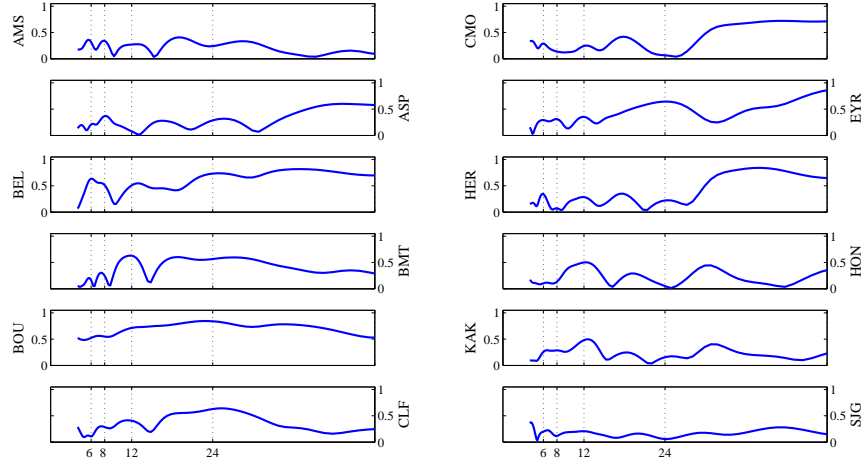}
  \caption{Modulus of the wavelet cross-correlation functions for each pair formed by VSS and one of the twelve chosen magnetic stations for June, 2007.}
  \label{fig:corrVSS}
\end{figure}

To exemplify the results obtained by calculating wavelet correlation, we selected the period of June, 2007.
Fig.~\ref{fig:corrVSS} shows the modulus of the correlations functions (for this period) for the twelve magnetic stations.
The correlation graphics are displayed by magnetic stations from top to bottom and divided in two blocks, in alphabetic order. 
These selected stations can be divided as follows: low latitude (ASP, BMT, HER, HON, KAK and SJG), medium latitude (AMS, BEL, BOU, CLF and EYR) and high latitude (CMO). 
On the vertical axis, we present the correlation coefficients and in the horizontal axis, the scale (period in hours).
The periods of 24, 12, 8 and 6-h are highlighted with dashed lines in order to facilitate visual inspection and comparison.
It is possible to verify that for most of the pairs $\mathcal{C}\left(a\right)$ varies considerably with scale.
We observe that the correlation between two geomagnetic data sets is scale dependent.
We also observe that the correlation is usually larger for the first harmonics of the diurnal variations, and it is smaller for the following harmonics.

The determination coefficients $D(a)$ were obtained by applying the equation~\ref{eq:det} to the values of coefficients extracted from the wavelet correlation curves (see Fig.~\ref{fig:corrVSS}).
Once we determined the correlation coefficients for 24 and 12-h periods, we were able to calculate the determination coefficients.
In order to facilitate the analysis of the determination coefficients, we calculated the mean determination coefficient $\overline{D(a)}$ for three different periods corresponding to high solar activity (years of 1999, 2000 and 2001), medium solar activity (years of 2002, 2003 and 2004) and low solar activity (years of 2005, 2006 and 2007).

\begin{figure}[htb]
    \begin{minipage}[h]{0.5\linewidth}
    \includegraphics[height=5cm,width=6.5cm]{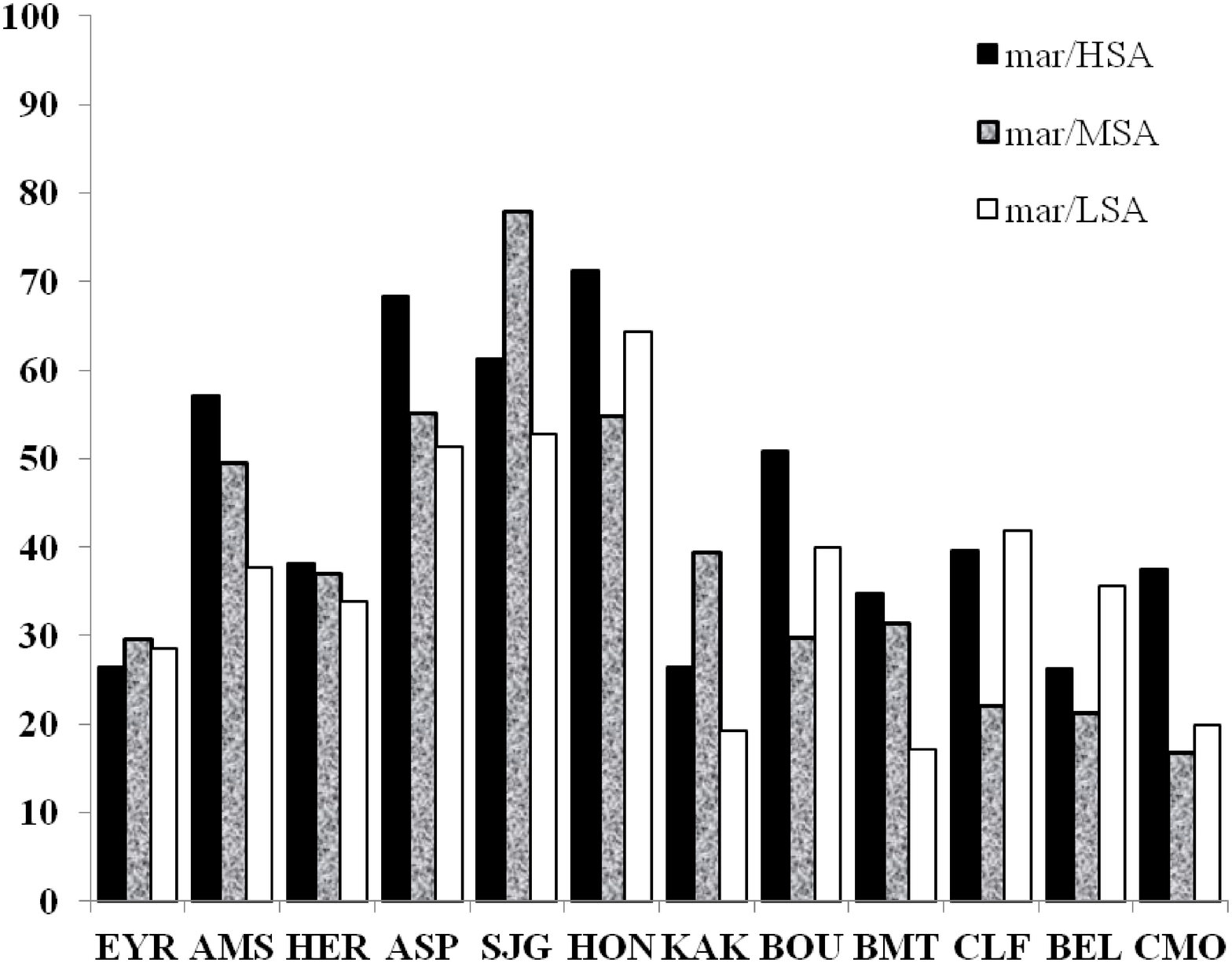}
\end{minipage}\hfill
 \begin{minipage}[h]{0.5\linewidth}
    \includegraphics[height=5cm,width=6.5cm]{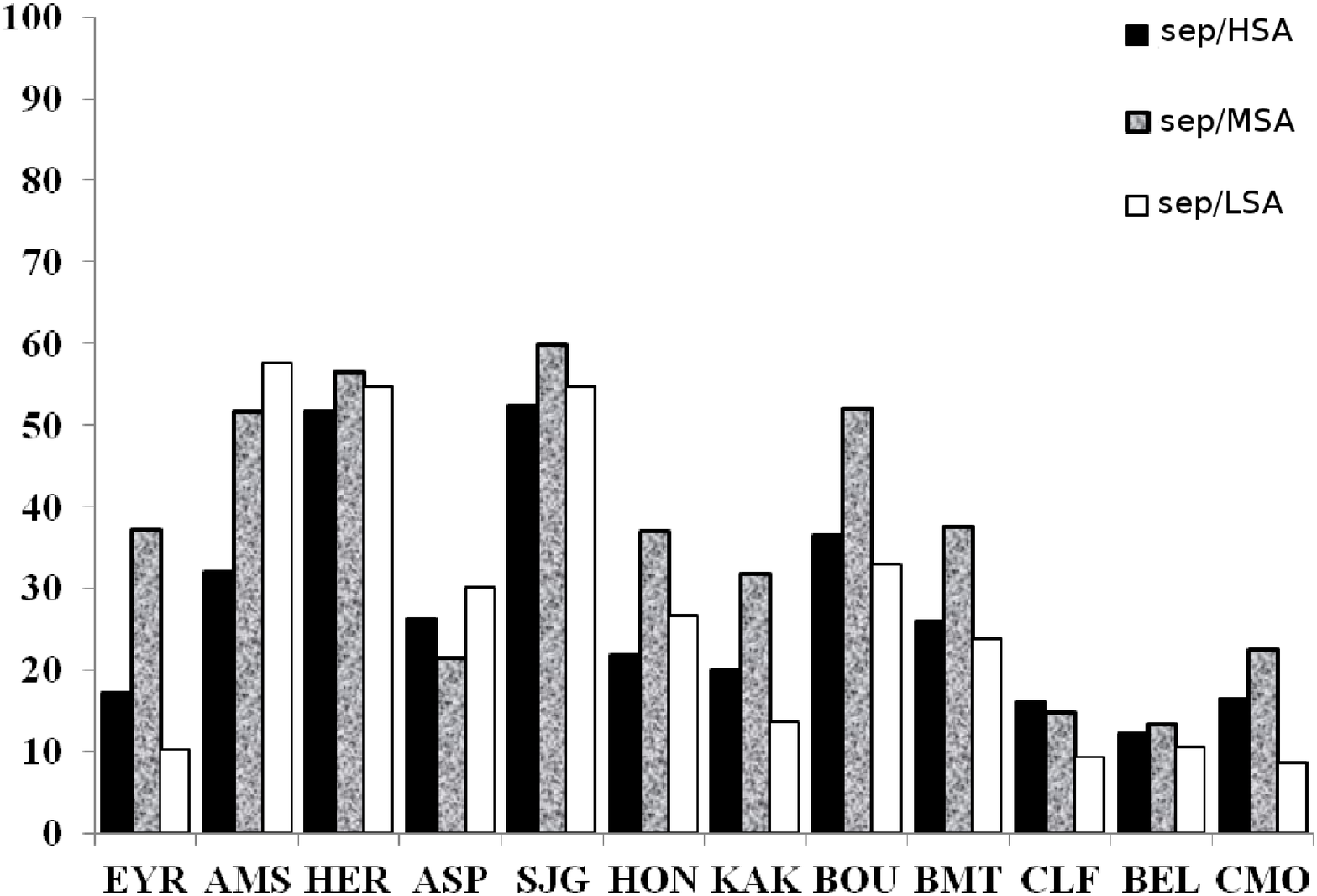}
\end{minipage}\\
\begin{minipage}[h]{0.5\linewidth}
    \includegraphics[height=5cm,width=6.5cm]{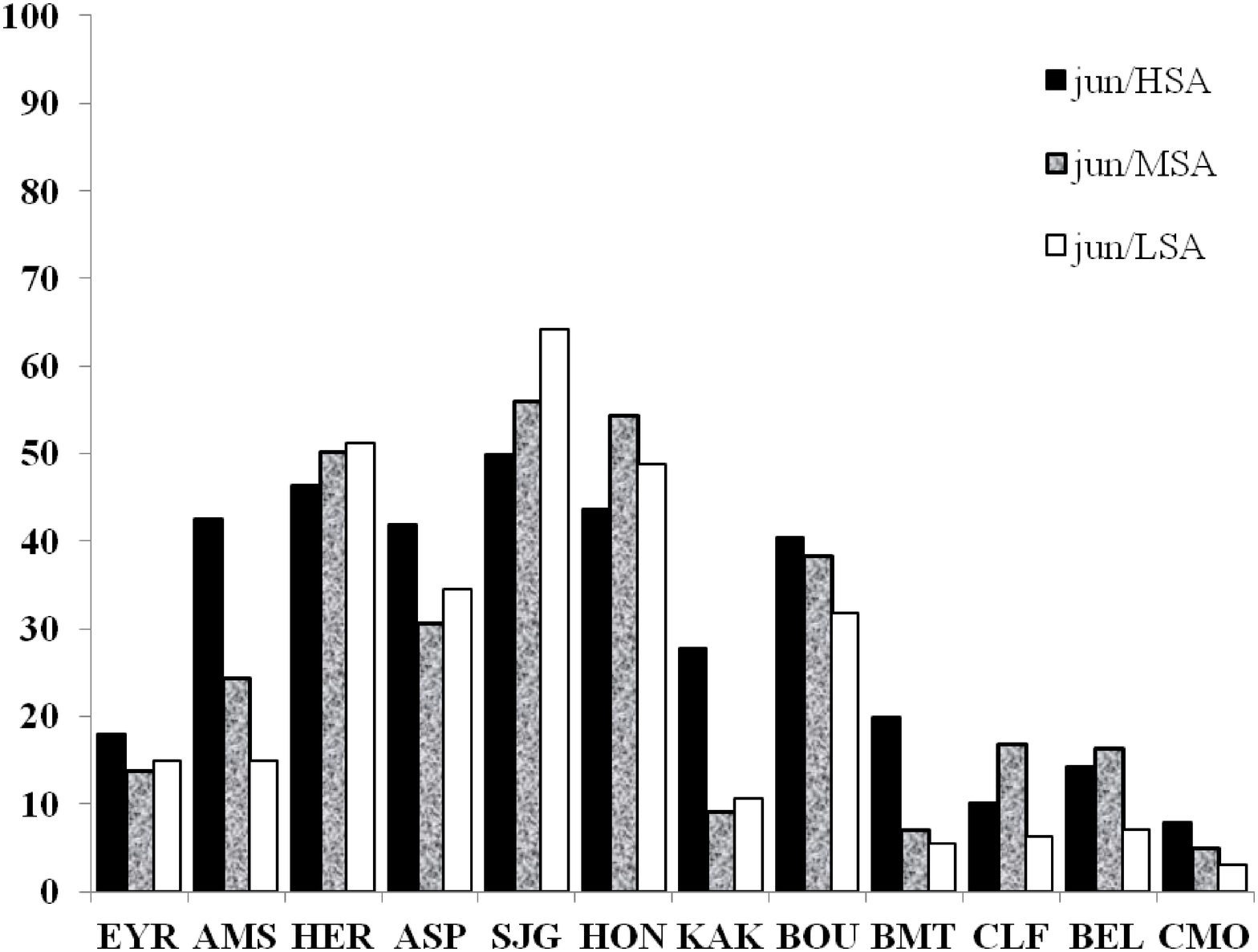}
\end{minipage}\hfill
 \begin{minipage}[h]{0.5\linewidth}
    \includegraphics[height=5cm,width=6.5cm]{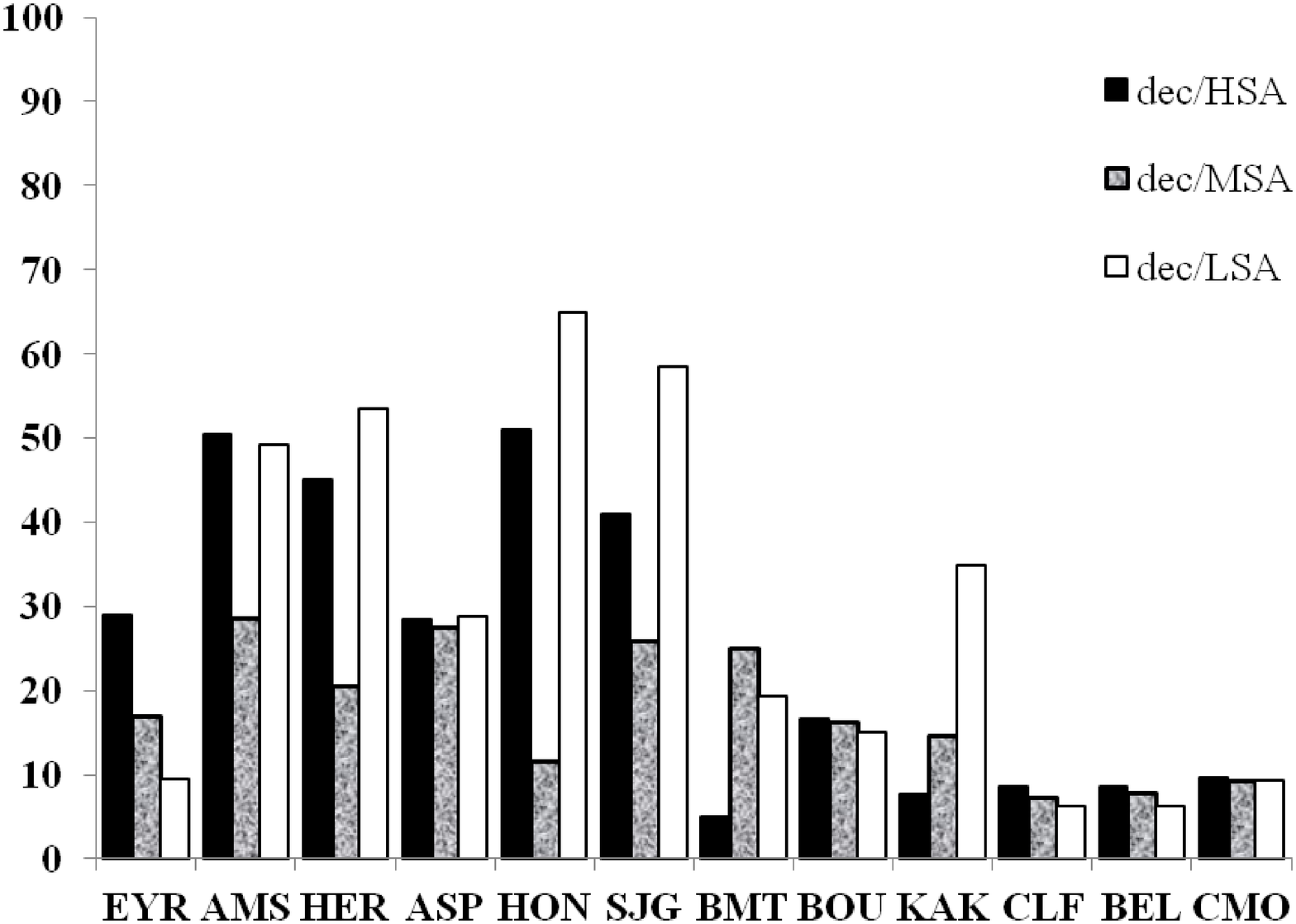}
\end{minipage}
\caption{Bar graphs of the diurnal geomagnetic variation during the equinoxes (March and September) and solstices (June and December) months for three different periods corresponding to high solar activity shown in black (years of 1999, 2000 and 2001), medium solar activity shown in grey (years of 2002, 2003 and 2004) and low solar activity shown in white (years of 2005, 2006 and 2007). The vertical axis shows the mean of determination coefficient $\overline{D(a)}$ and the horizontal shows the IAGA code of the magnetic stations distributed from the lowest latitude on the South Hemisphere to the highest latitude on the North Hemisphere.}
\label{fig:24h}
\end{figure}

\begin{figure}[htb]
    \begin{minipage}[h]{0.5\linewidth}
    \includegraphics[height=5cm,width=6.5cm]{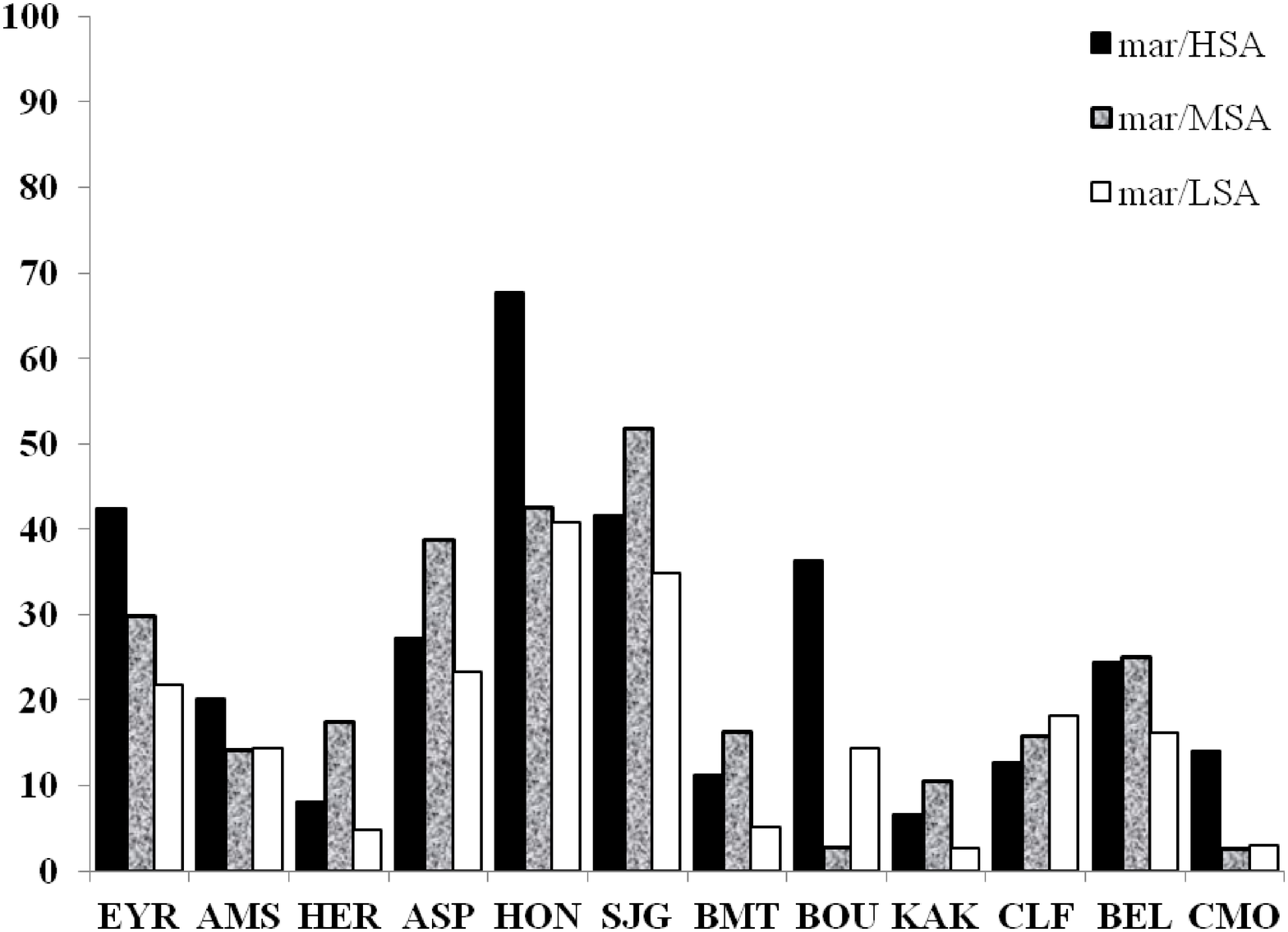}
\end{minipage}\hfill
 \begin{minipage}[h]{0.5\linewidth}
    \includegraphics[height=5cm,width=6.5cm]{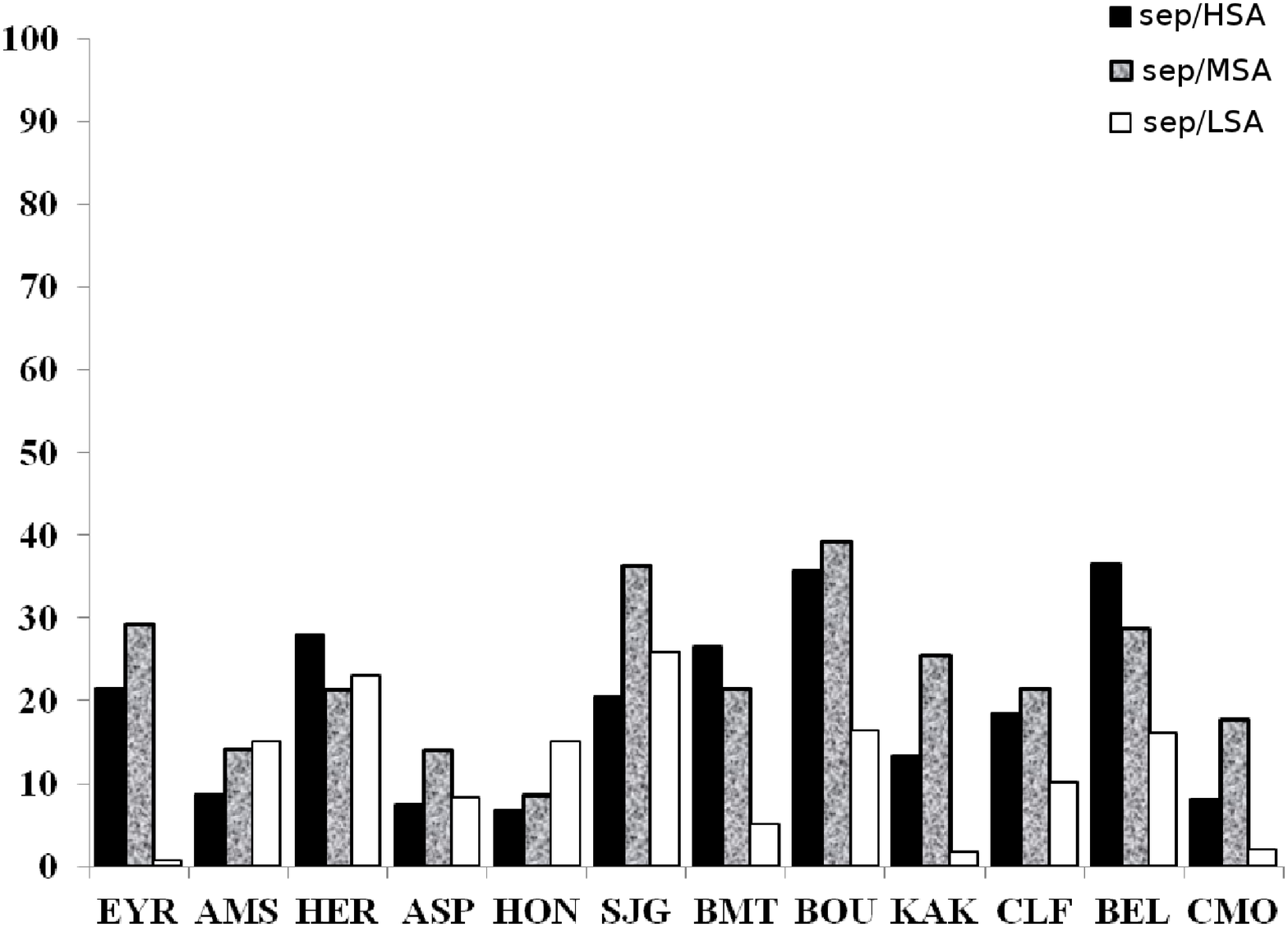}
\end{minipage}\\
\begin{minipage}[h]{0.5\linewidth}
    \includegraphics[height=5cm,width=6.5cm]{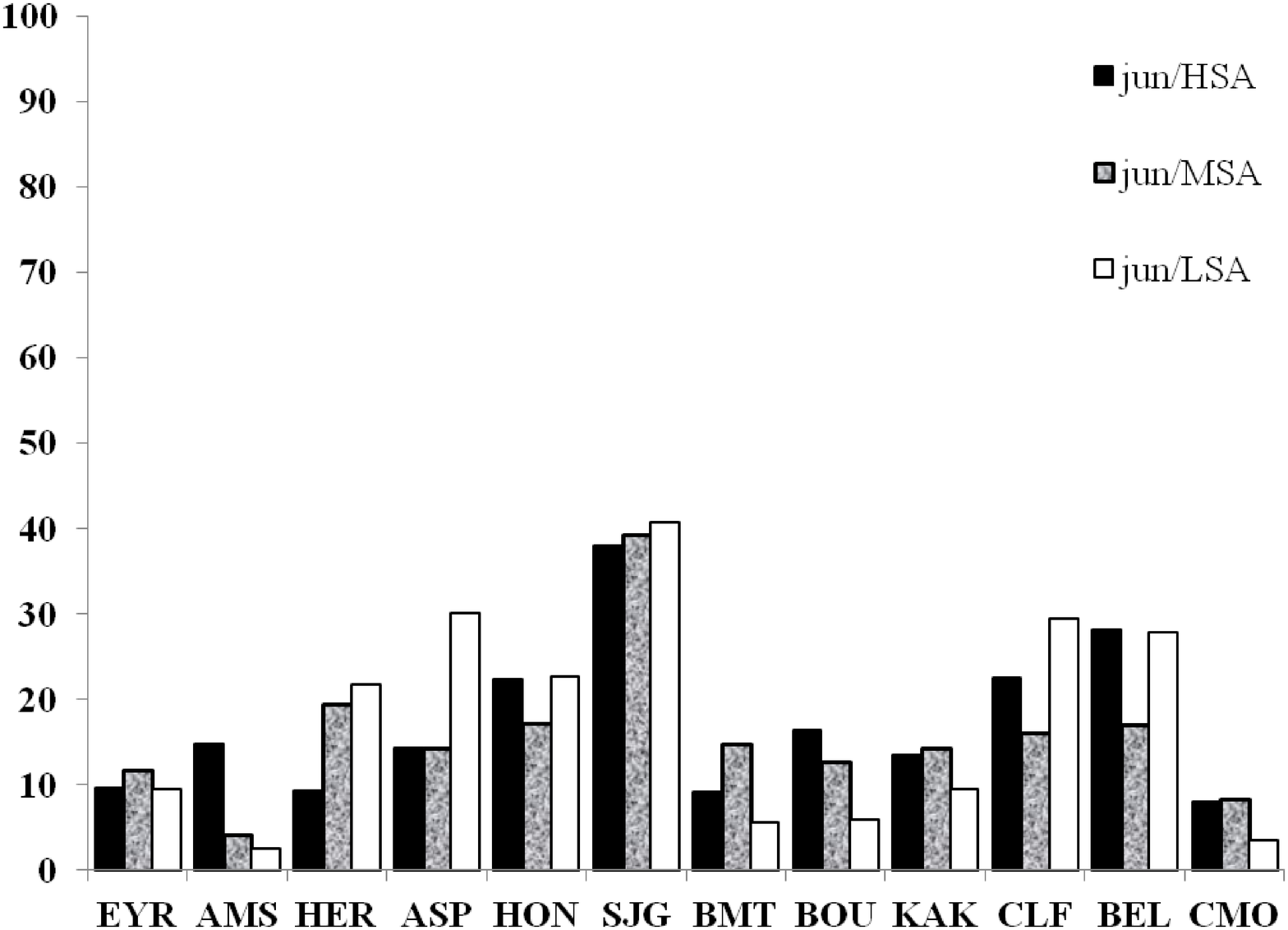}
\end{minipage}\hfill
 \begin{minipage}[h]{0.5\linewidth}
    \includegraphics[height=5cm,width=6.5cm]{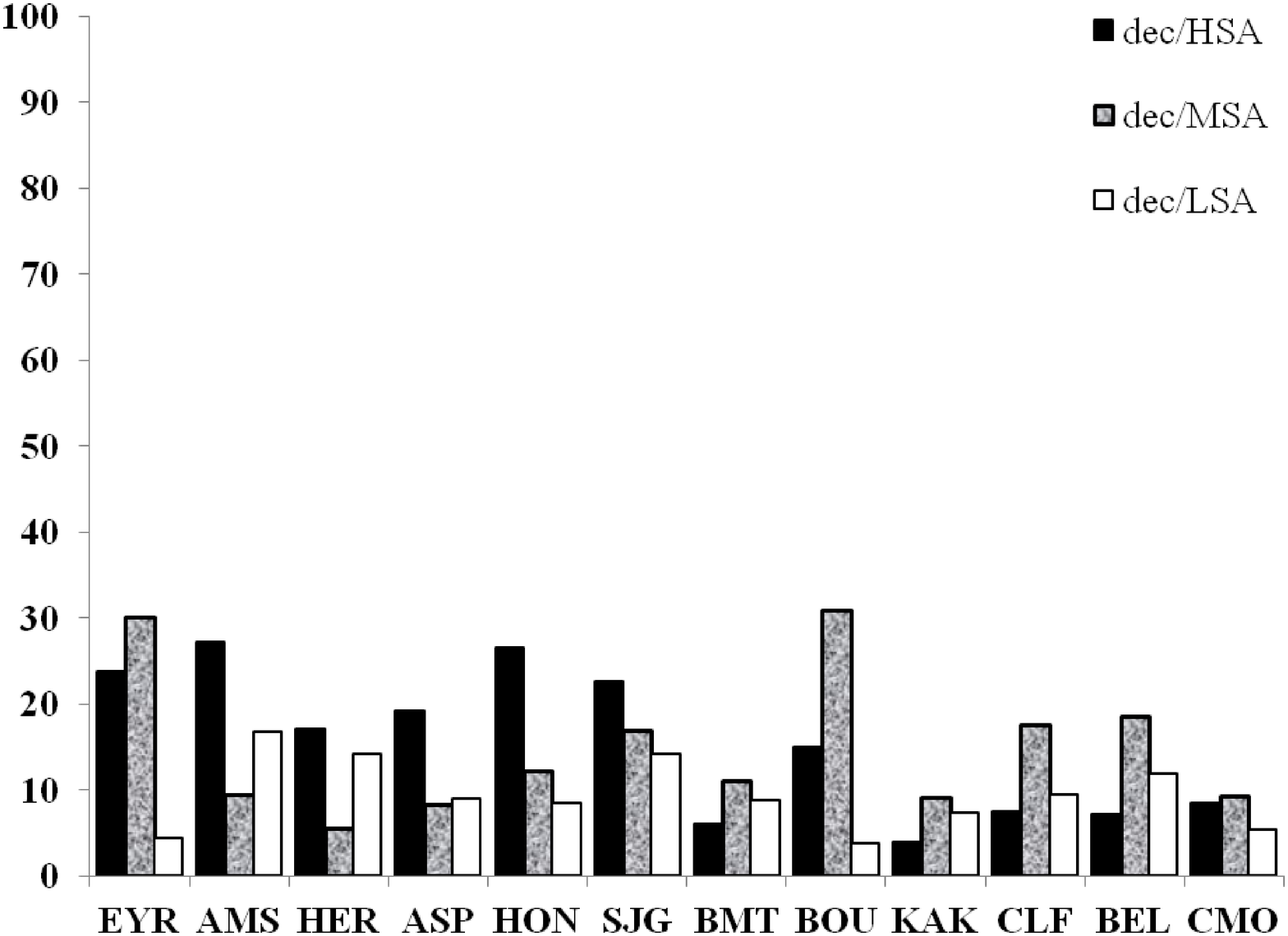}
\end{minipage}
\caption{Bar graphs of the semidiurnal geomagnetic variation during the equinoxes (March and September) and solstices (June and December) months for three different periods corresponding to high solar activity shown in black (years of 1999, 2000 and 2001), medium solar activity shown in grey (years of 2002, 2003 and 2004) and low solar activity shown in white (years of 2005, 2006 and 2007). The vertical axis shows the mean of determination coefficient $\overline{D(a)}$ and the horizontal shows the IAGA code of the magnetic stations distributed from the lowest latitude on the South Hemisphere to the highest latitude on the North Hemisphere.}
\label{fig:12h}
\end{figure}

Fig.~\ref{fig:24h} and \ref{fig:12h} show the mean determination coefficient for the 12 chosen stations during the equinoxes and solstices between the years of 1999 and 2007. 
These figures present the diurnal and semidiurnal geomagnetic variations, respectively, and their seasonal behavior.
In the vertical axis, we present the determination coefficient and in the horizontal axis, the twelve chosen magnetic stations ordered by latitude from the Southern to the Northern Hemisphere.
This distribution of stations helps to verify the latitudinal dependence of the diurnal variations.
The nine years interval enables the study of the solar activity dependence upon the Sq variations. 
Years of high solar activity are shown in black (corresponding to the mean value of the determination coefficient for the years of 1999, 2000 and 2001), years of medium solar activity, in grey (corresponding to the mean value of the determination coefficient for the years of 2002, 2003 and 2004) and years of low solar activity, in white (corresponding to the mean value of the determination coefficient for the years of 2005, 2006 and 2007).

In Fig.~\ref{fig:24h}, it is possible to identify a primary latitudinal dependence for the equinoxes and solstices.
At equinoxes, the stations of AMS, ASP and HON have the largest values of the mean determination coefficients considering the three periods of analysis for March, and the stations of AMS, HER and SJG, for September.
If we consider the simple dipole representation of the Earth's main magnetic field and the north-south symmetry between Hemispheres of thermospheric wind and conductivity field in equinox conditions, we would find that the stations at the same latitude should have the same determination coefficient.
The stations of EYR (located at Geomagnetic coord. Lat. $46.79^o$) and AMS (located at Geomagnetic coord. Lat. $46.07^o$), and of SJG (located at Geomagnetic coord. Lat. $27.93^o$) and KAK (located at Geomagnetic coord. Lat. $27.46^o$) should have similar values of the mean determination coefficients at the equinoxes.
By analyzing Fig.~\ref{fig:24h}, this fact does not hold for the months of March and September. 
The geomagnetic field is asymmetric, thus, the mean conductance for the Northern and Southern Hemisphere will have different values even at equinox \citep{Stening1971}.
This characteristic may explain the difference of response between the magnetic stations of EYR and AMS because they are located at similar magnetic latitudes but at different Hemispheres.
However, this does not explain the difference of response between the magnetic stations of SJG and KAK.
Also, the stations of EYR, AMS, ASP, HON, SJG, BMT, KAK and BEL have largest values of the mean determination coefficients during the high and medium solar activity than during the low solar activity on March, and the stations of EYR, BMT, BOU, KAK, CLF, BEL and CMO, on September.

The disparity in the response of magnetic stations between solstices is expected.
In Fig.~\ref{fig:24h}, we expect low values of the mean determination coefficients between VSS and the magnetic stations located in the Northern Hemisphere (BEL, BMT, BOU, CMO, HON, KAK and SJG) as a result of the month of June and December be solstices, and consequently, of the wind asymmetry between the Northern and Southern Hemispheres due to the contrast in summer-winter radiation. 
Even so, latitudinal dependence prevailed at the solstices.
The stations of HER, ASP, HON, SJG, CLF and BEL have the largest values of the mean determination coefficients considering the three periods of analysis for June.
The stations HON, SJG, CLF and BEL located on the Northern Hemisphere presented larger mean determination coefficients which can be attributable to the latitudinal dependence prevalence of the Sq current over the wind asymmetry. 
The largest values for December are not as perceptible as for the other months.
Also, the stations of HER, ASP and SJG have largest values of the mean determination coefficients during the medium and low solar activity than during the high solar activity on June, and the stations of BMT, KAK, CLF and BEL, on December.
This characteristic may be explained by the thermosphere wind asymmetry during the minimum solar activity which is not as larger as during the maximum solar activity. 
Magnetic stations of HER, HON and SJG located at lower latitudes as VSS usually present larger mean determination coefficients.
These differences in responses between the stations located at similar latitudes as EYR and AMS, HER and ASP, SJG and KAK and BOU and CLF can be explained by the following features of the Sq field reported by \Citet{MatsushitaMaeda1965}: 1) the total Sq current intensity is about 1.5 times larger in the Hemisphere's summer than in the winter and 2) the latitudinal position of the external current center is higher in summer than in winter.

In this work, we also find that the 24-h diurnal variation presented differences values of the mean determination coefficient for magnetic stations located in approximately the same geomagnetic latitude (EYR and AMS, HER and ASP, SJG and KAK, and BOU and CLF) but different longitudes.
This means that these stations are affected differently by the E-layer dynamo. 
At equinoxes, the mean determination coefficient was usually larger in years of higher solar activity, and at solstices it is practically larger in the year of lower solar activity. 
These results may be explained by the variation of conductivity of the E-region which has an ionization density dependency \citep{ForbesLindzen1976}. 
Also, the total geomagnetic field affects the conductivity through its dependence on electron and ions gyro-frequencies \citep{Stening1971}.

Another strong influence, very important to be referred, is the intensity of the geomagnetic field.
The total geomagnetic field is particularly high in the regions of Central Canada, Siberia and South of Australia and it is quite low near Southern Brazil \citep{Campbell1997}. 
In Fig.~\ref{fig:24h}, it is possible to observe that the magnetic stations, located at the Northwest of Canada (CMO), North of the United States (BOU), China (BMT), Japan (KAK) and Central Australia (ASP), where the geomagnetic field is particularly high, presented lower determination coefficient when compared to other stations.

For the semi-diurnal variation, the latitudinal dependence is not as perceptible as in the diurnal variation (see Fig.~\ref{fig:24h} and \ref{fig:12h}). 
Also, the equinoxes and solstices present irregular distribution due to the solar activity.

A lower correlation between VSS and CMO is expected because CMO is a station located at high latitude. 
In this region, the influence of the Sq currents is limited when compared to low and mid-latitudes.
At high latitudes, the major influence comes from a large horizontal current that flows in the D and E regions of the auroral ionosphere, called the Auroral Electrojet. 
There are two characteristic of the auroral region that we must mentioned.
First of all, the conductivity in the auroral ionosphere is generally larger than at lower latitudes.
Second, the horizontal electric field in the auroral ionosphere is also larger than at lower latitudes. 
Since the strength of the current flow is directly proportional to the vector product of the conductivity and the horizontal electric field, the auroral electrojet currents are generally larger if compared to those at lower latitudes (see \Citet{Sizova2002} for more details).

\section{Conclusions}
\label{Summary}

A world-wide distribution of the harmonic components of solar diurnal variations (24 and 12-h periods) has been observed in latitude and longitude by using the gapped wavelet analysis and the wavelet cross-correlation technique.
The objective of this paper is to study the characteristics of these variations at a Brazilian station as compared to the features from other magnetic stations to better understand the dynamics of the diurnal variations involved in the monitoring of the Earth's magnetic field.
The main results in this analysis can be summarized as follows:

\begin{enumerate}
 \item[1.] The stations of EYR (Geomagnetic coord. Lat. $46.79^o$) and AMS (Geomagnetic coord. Lat. $46.07^o$); and of SJG (Geomagnetic coord. Lat. $27.93^o$) and KAK (Geomagnetic coord. Lat. $27.46^o$) located at similar latitudes presented very different values of the mean determination coefficients at the equinoxes. This characteristic may be explained by  the difference in the mean conductance for the Northern and Southern Hemisphere due to the asymmetry of the geomagnetic field.
\end{enumerate}

\begin{enumerate}
 \item[2.] At equinoxes and solstices, magnetic stations located at the Northwest of Canada (CMO), North of the United States (BOU), China (BMT), Japan (KAK) and Central Australia (ASP), where the geomagnetic field is particularly high, presented lower mean determination coefficient when compared to other stations. This fact can be explained by the peculiar location of the VSS, under the South Atlantic Magnetic Anomaly (minimum geomagnetic field intensity).
\end{enumerate}

\begin{enumerate}
 \item[3.] For the diurnal variation, the determination coefficient was usually larger in years of higher solar activity at the equinoxes. The dependence of the E-region conductivity on the ionization density may explain the variation of the mean determination coefficient due to the solar activity.
\end{enumerate}

\begin{enumerate}
 \item[4.] The 24-h diurnal variation presented different values of the mean determination coefficient for magnetic stations located in approximately the same geomagnetic latitude (EYR and AMS; HER and ASP; SJG and KAK; and BOU and CLF) but different longitudes. The anisotropy of the E-region electric conductivity because of the effect of the geomagnetic field also may account for the differences of the mean determination coefficient among these stations.
\end{enumerate}

\begin{enumerate}
 \item[5.] The latitudinal dependence for the semi-diurnal variation was not as noticeable as for the diurnal variation, and also, presented an irregular distribution due to the solar activity.
\end{enumerate}

\begin{enumerate}
 \item[6.] In closing, this analysis has shown a great spatial and temporal variability of the diurnal and semidiurnal variations, and also, unequal contributions for each station according to their correlation to VSS due to the conductivity of the E-region, the geomagnetic field intensity and its configuration, and thermospheric winds.
\end{enumerate}

The gapped wavelet analysis is an alternative technique to study data series that contain gaps of various sizes.
Consequently, as proposed here, it can be used to exclude disturbed days and to study the characteristics of solar diurnal variations.
Using a network of magnetic stations, we determine the contributions of the longitude and latitude and also of the total geomagnetic field. 


\section{Acknowledgments}
V. Klausner wishes to thanks CAPES for the financial support of her PhD (CAPES -- grants 465/2008) and her Postdoctoral research (FAPESP -- 2011/20588-7).
This work was supported by CNPq (grants 309017/2007-6, 486165/2006-0, 308680/2007-3, 478707/2003, 477819/2003-6, 382465/01-6), FAPESP (grants
2007/07723-7) and CAPES (grants 86/2010-29, 0880/08-6, 86/2010-29, 551006/2011-0, 17002/2012-8). 
Also, the authors would like to thank the INTERMAGNET programme for the datasets used in this work. 
The authors sincerely acknowledge a reviewer whose comments and advices have greatly contributed to improve the final form of this work


%
%


\bibliographystyle{abbrv}

\appendix

\section{}
\label{Appendice A}

\begin{figure}[htb]
\centering
\begin{tabular}{cc}
a) Morlet with $\sigma=0.25$\\
 \includegraphics[width=0.7\textwidth]{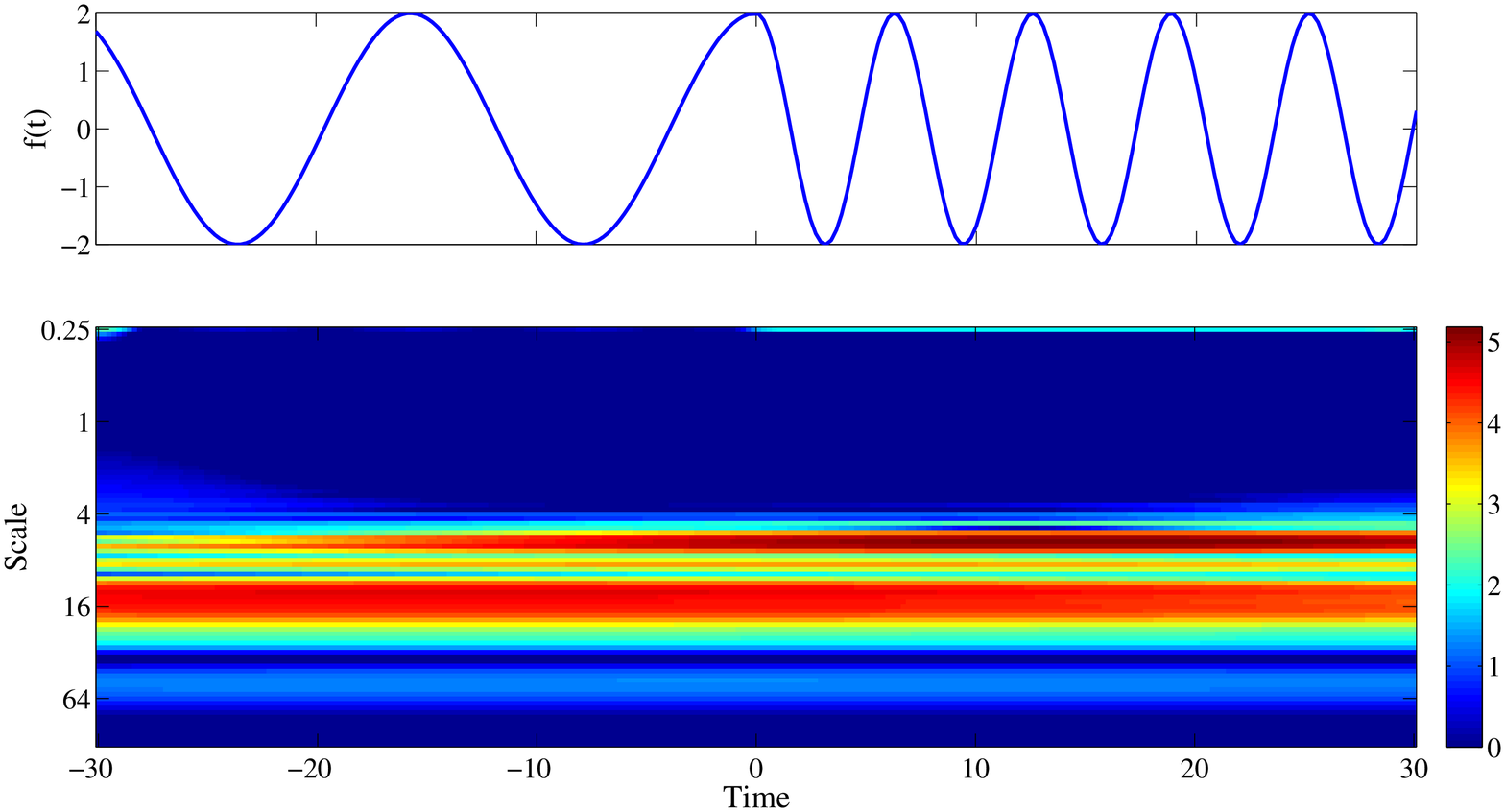}\\
b) Morlet with $\sigma=1$\\
 \includegraphics[width=0.7\textwidth]{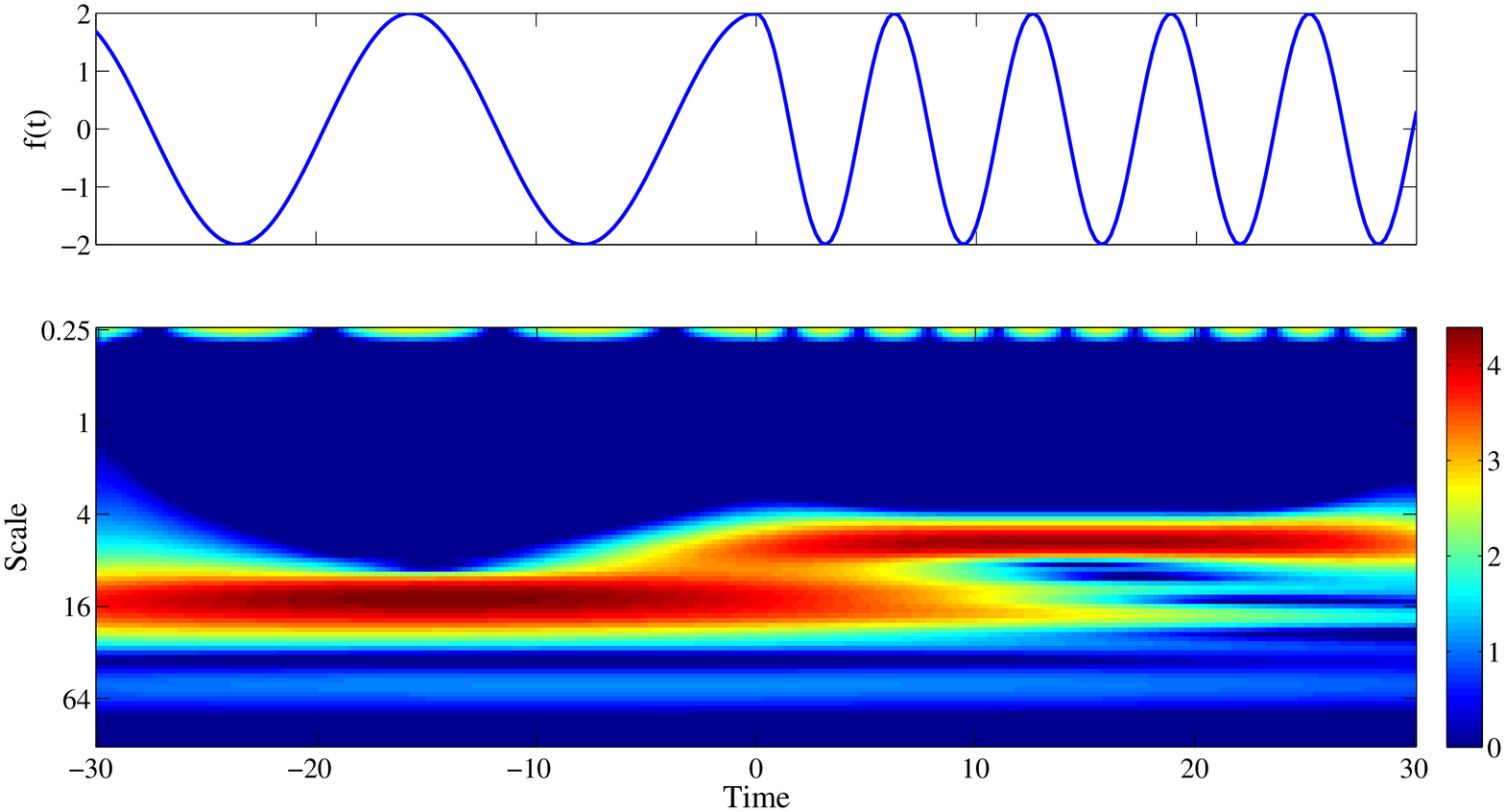}\\
c) Morlet with $\sigma=4$\\
\includegraphics[width=0.7\textwidth]{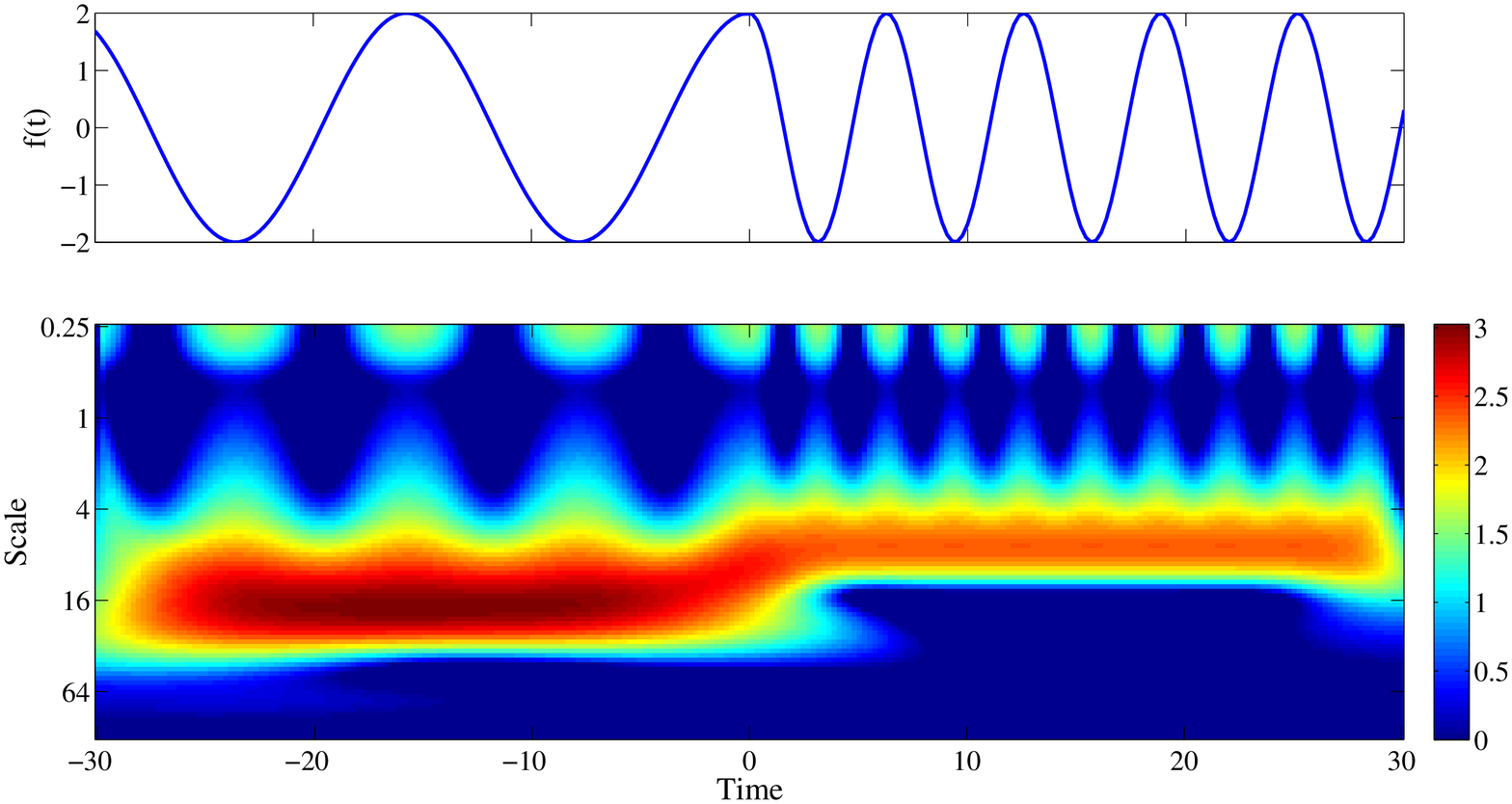}\\
\end{tabular}
\caption{Representations of an oscillatory function with low and high frequencies spatially displaced and the respective wavelet spectrum for Morlet with different values of parameter$\sigma$. The parameter $\sigma$ assumed the following values of (a) equal to $0.25,\,\text{(b)}\;1$ and (c) $4$.}
\label{fig:abruptfreq}
\end{figure}

In order to demonstrate the possibilities of the use of different parameters, and its dependence on the changing characteristics of the chosen wavelet, a sinusoidal signal containing two different frequencies is analyzed by means of the Morlet wavelet in Fig.~\ref{fig:abruptfreq}.
This figure contains $3$ panels.
Each panel shows the signal representation and the modulus scalogram using the Morlet wavelet.
It is possible to observe that there is a better frequency localization of the Morlet wavelet transform as the parameter $\sigma$ increases, see Fig.~\ref{fig:abruptfreq}(a), although it has a worse time localization.
The Fig.~\ref{fig:abruptfreq}(b) occurs the opposite of Fig.~\ref{fig:abruptfreq}(a), it has a better time localization and worse frequency localization.

In Fig.~\ref{fig:abruptfreq}(c), it is also possible to observe that the lower values of the wavelet coefficients indicate a transition region between different types of movements.

We adopted $\omega_{0}=\,6$ and $\sigma=1$ that also provides the better possible time-frequency equilibrium, see Fig.~\ref{fig:abruptfreq}(b).
In this case also, the Morlet wavelet is suited very well for experimental data analysis because it has a Gaussian envelope.
Thus, it allows reaching a reasonable compromise between time and frequency resolutions.

\section{}
\label{Appendice B}

In this section, an analysis of a synthetic signal is performed to illustrate the results and the interpretation of the CWT and the CWT using gapped wavelet techniques, with the purpose to highlight the differences and the advantages as mentioned earlier.
In order to exemplify the wavelet scalogram expected from this analysis, we defined a synthetic Sq to reproduce the daily magnetic variation, similarly to the described in \Citet{Maslovaetall2010},

\begin{equation}
  S_{q} \left(t\right)=\sin\left(\frac{2 \pi t}{24}\right)+\sin\left(\frac{4\pi t}{24}\right),
\end{equation}

and the noise $R(t)$ as

\begin{equation}
R\left(t\right)=0.5*V\left(t\right)*\sin\left(\frac{2 \pi t}{24}\right).
\end{equation}

The good Sq estimative $Q(t)$ is given by

\begin{equation}
 Q(t)=[S_{q}(t)+R(t)]*U(t),
\end{equation}
where the day to day variability was introduced by the noise variables $V(t)$, $U(t)$ and $R(t)$, where $V(t)$ was uniformly distributed on $[0,1]$ and $U(t)$ on $[0.5,1.5]$.

\begin{figure}[htb]
    \includegraphics[width=0.9\textwidth]{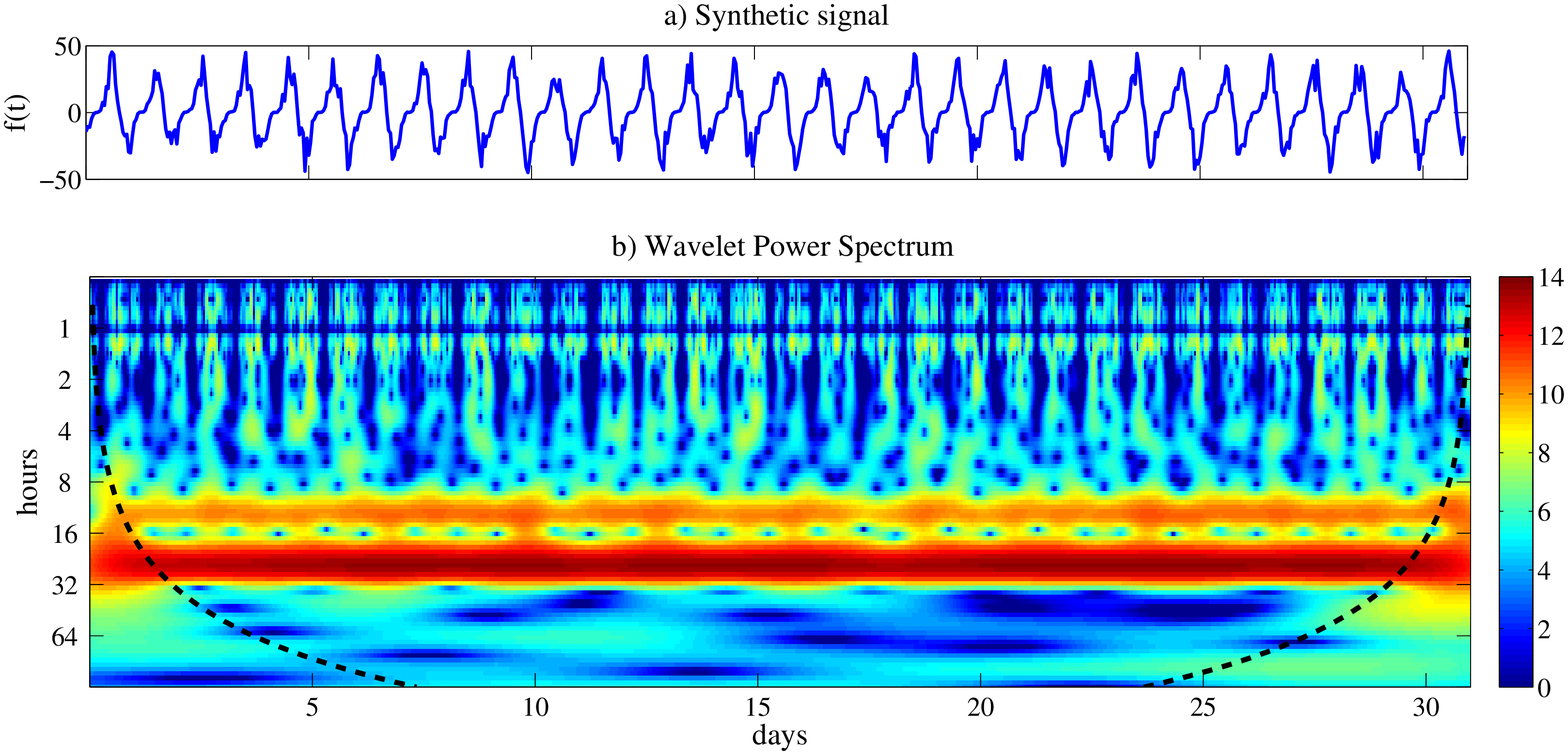}\\
\caption{Spectrograms of synthetic signal consisting of daily magnetic variations. From top to bottom, each panel shows: (a) Synthetic signal and (b) the local wavelet power spectrum using Morlet wavelet, logarithmic scaled representing $\log2{(|W(a,b)|)}$. }
\label{fig:1a}
\end{figure}

Fig.~\ref{fig:1a}, from top to bottom, shows the synthetic geomagnetic signal and the wavelet square modulus (scalogram) with Morlet wavelet. 
It was possible to identify the two most prevailing periods (12 and 24-h) due to the quiet variations. 
Also, we can identify the change of frequencies introduced by the noise, which indicates that the wavelet analysis is useful for these multi-scale phenomena. 

\begin{figure}[htb]
    \includegraphics[width=0.9\textwidth]{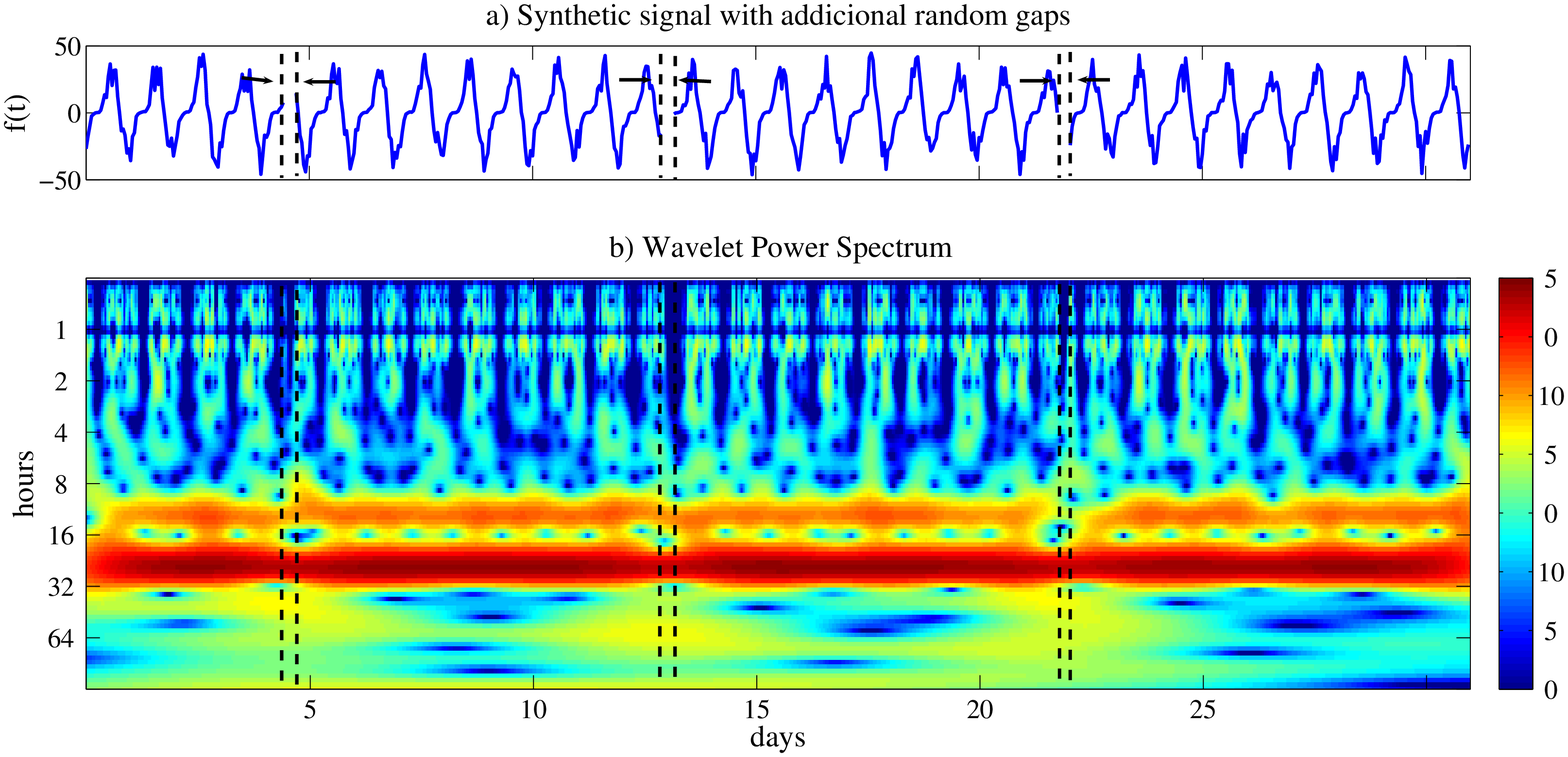}\\
\caption{Spectrograms of synthetic signal with additional random gaps. From top to bottom, each spectrogram shows: (a) Synthetic signal with additional random gaps and (b) the local wavelet power spectrum using gapped wavelet representing $\log2{(|W(a,b)|)}$. }
\label{fig:1b}
\end{figure}

In Fig.~\ref{fig:1b}, we used the same synthetic signal used in Fig.~\ref{fig:1a} but modified with additional random gaps in the signal.
The areas with gaps were signalized by arrows and dashed lines.
We applied the gapped wavelet in this signal.
In this new analysis, it was also possible to identify the quiet geomagnetic variations with periods of 12 and 24-h and there not a significantly difference in the common intervals of the CWT and gapped scalograms, Figs.~\ref{fig:1a} and \ref{fig:1b}, respectively. 
However, there was a little effect in the boundaries surrounding the gaps regions which led to some energy lost.
This energy lost was not relevant, as seen in the comparison, see Figs.~\ref{fig:1a} and \ref{fig:1b}, and can be neglected.

In an attempt to fill the gaps, we first tried to use cubic splines interpolation, which led to an underestimation of high frequencies.
As the signals may have many spectral components, only by using gapped wavelet, the high-frequency part of the spectrum can be reconstructed.
For this reason, the gapped wavelet has some advantages over interpolation.
However, the gapped wavelet has some limitations, it can only be used if the gap is not larger than the period of analysis. 

After having applied the CWT and gapped wavelets in the synthetic signal, as presented in Figs.~\ref{fig:1a} and \ref{fig:1b}, the conclusion was that the gapped wavelet fulfilled better the purposes of this work.

\bibliographystyle{elsarticle-harv}





\end{document}